\documentclass[journal]{IEEEtran}
\usepackage{algorithmicx}
\usepackage[ruled,vlined,linesnumbered]{algorithm2e}
\usepackage{hhline}
\usepackage{amsmath,mathtools}
\usepackage{amsfonts,amssymb}
\usepackage{mathrsfs}
\usepackage{gensymb} % for degree
\usepackage{caption}% http://ctan.org/pkg/caption
\usepackage{multirow}
\usepackage{graphicx} 
\usepackage{multirow}
\let\labelindent\relax
\usepackage{enumitem,color}
\usepackage{algpseudocode}
\begin{document}
\title{Real-Time Dynamic Map with Crowdsourcing Vehicles in Edge Computing}
  
\author{
        Qiang Liu,~\IEEEmembership{Member,~IEEE,}
        Tao Han,~\IEEEmembership{Senior Member,~IEEE,}\\
        Jiang (Linda) Xie, ~\IEEEmembership{Fellow,~IEEE,}
        and BaekGyu Kim, ~\IEEEmembership{Member,~IEEE,}\vspace{-0.05in}% <-this % stops a space
\IEEEcompsocitemizethanks{
        \IEEEcompsocthanksitem Qiang Liu is with the School of Computing, University of Nebraska-Lincoln.
        E-mail: qiang.liu@unl.edu
        \IEEEcompsocthanksitem Tao Han is with the Department of Electrical and Computer Engineering, New Jersey Institute of Technology.
        E-mail: tao.han@njit.edu
        \IEEEcompsocthanksitem Jiang (Linda) Xie is with the Department of Electrical and Computer Engineering, University of North Carolina at Charlotte.
        E-mail: linda.xie@uncc.edu
        \IEEEcompsocthanksitem BaekGyu Kim is with the Department of Information and Communication Engineering, Daegu Gyeongbuk Institute of Science and Technology.
        E-mail: bkim@dgist.ac.kr
}% <-this % stops an unwanted space
\thanks{Partial contents of this article appeared in IEEE International Conference on Computer Communications 2021~\cite{liu2021livemap}. }}

% \markboth{IEEE TRANSACTIONS ON MICROWAVE THEORY AND TECHNIQUES, VOL.~60, NO.~12, DECEMBER~2012
% }{Roberg \MakeLowercase{\textit{et al.}}: High-Efficiency Diode and Transistor Rectifiers}

% ====================================================================
\maketitle

% === ABSTRACT ====================================================================
% =================================================================================
\begin{abstract}
Autonomous driving perceives surroundings with line-of-sight sensors that are compromised under environmental uncertainties. To achieve real time global information in high definition map, we investigate to share perception information among connected and automated vehicles. However, it is challenging to achieve real time perception sharing under varying network dynamics in automotive edge computing. In this paper, we propose a novel real time dynamic map, named \emph{LiveMap} to detect, match, and track objects on the road.
We design the data plane of \emph{LiveMap} to efficiently process individual vehicle data with multiple sequential computation components, including detection, projection, extraction, matching and combination. We design the control plane of \emph{LiveMap} to achieve adaptive vehicular offloading with two new algorithms (central and distributed) to balance the latency and coverage performance based on deep reinforcement learning techniques. We conduct extensive evaluation through both realistic experiments on a small-scale physical testbed and network simulations on an edge network simulator.
The results suggest that \emph{LiveMap} significantly outperforms existing solutions in terms of latency, coverage, and accuracy.
\end{abstract}

% === KEYWORDS ====================================================================
% =================================================================================
\begin{IEEEkeywords}
Dynamic Map, Edge Computing, Autonomous Driving
\end{IEEEkeywords}

% \fi
\IEEEpeerreviewmaketitle

% ====================================================================
% ====================================================================
% ====================================================================

% === I. INTRODUCTION =============================================================
% =================================================================================
\vspace{-0.1in}
\section{Introduction}
\label{sec:introduction}
\IEEEPARstart{A}{utonomous} driving and advanced driving assistance system (ADAS) are being evolved with the development of modern machine learning and pervasive parallel computing. 
Vehicles leverage a variety of sensors, e.g., camera and LiDAR, to perceive surroundings, and use onboard computers to understand the collected raw data in real time, e.g., semantic segmentation and object recognition.
With the high-definition (HD) map, advanced vehicular control algorithms accurately relocalize the vehicle and can tackle road situations with the perceived environmental context, e.g., pedestrians and lanes.

%%%%-----------  LOS limitation, HD map, -------------%%%%
Achieving highly reliable and safe driving, however, is very challenging, based on non-real-time HD map and individual vehicle perception.
On the one hand, the HD map~\cite{jiao2018machine}, including geometric, semantic, and map-prior layer, has no real time road information, e.g., pedestrian and vehicles, in the time scale of subseconds.
On the other hand, the perceptions of individual vehicles are limited and might be compromised under a variety of environmental uncertainties such as weather and occlusion~\cite{yaqoob2019autonomous}.
For example, existing line-of-sight vehicle sensors are with limited sensing ranges, which indicates that they cannot perceive information in occluded areas~\cite{qiu2018avr}.
Considering a car follows a truck that blocks the car's front sensor, passing the truck without the information about the opposite lane is unsafe.
% Besides, the perception range and fidelity of sensors might be further impaired by extreme weather, e.g., rain, snow, and dust.

% Thus, relying on sensors in a single vehicle alone may not be sufficient to fulfill high-safety driving.
%that are incapable of providing consistent and high-fidelity information of the environment,
% Lack of consistent and confidential information of surrounding environment poses a huge challenge toward full autonomous driving.   
\begin{figure}[!t]
	\centering
	\includegraphics[width=3in]{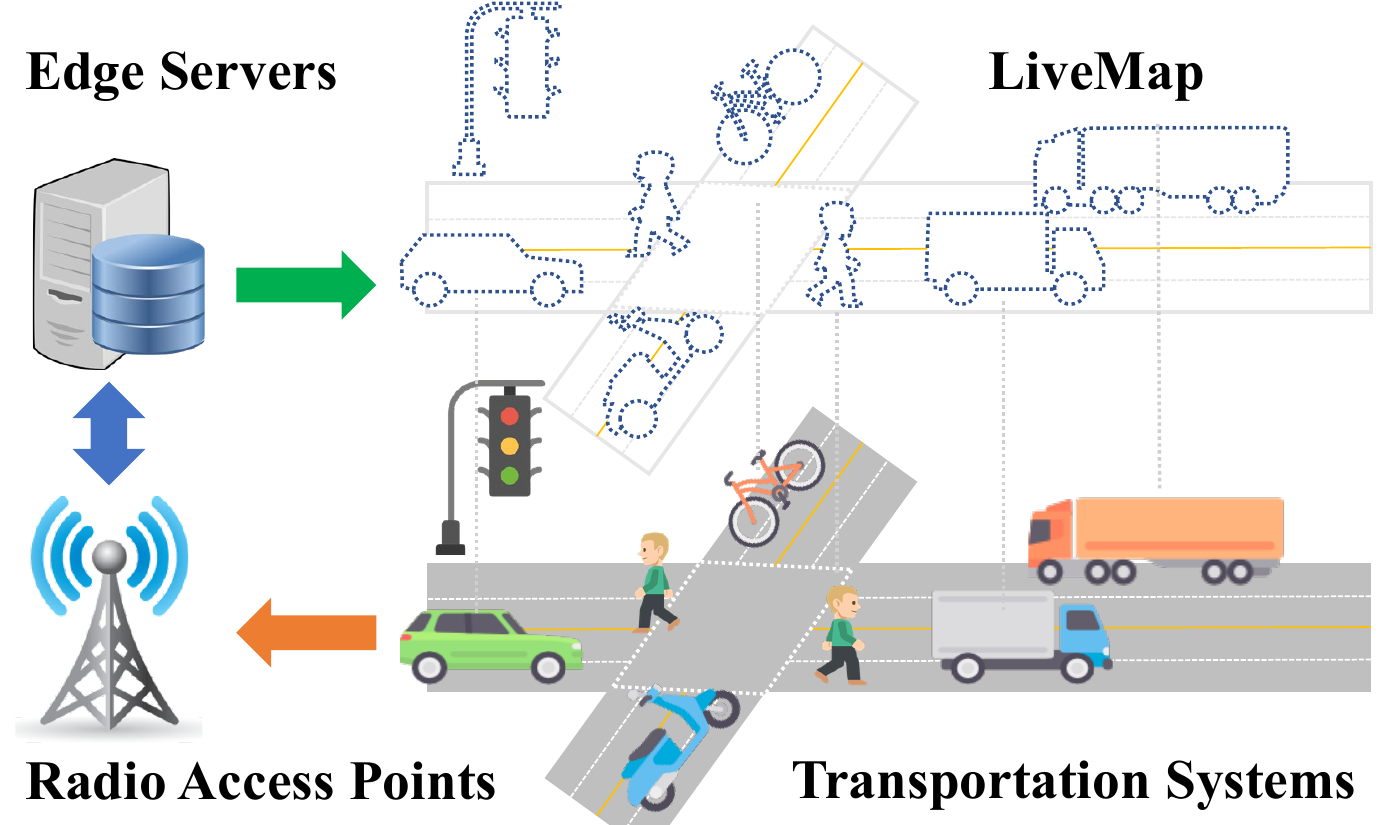}
	\vspace{-0.05in} \caption{\small An example of automotive edge computing.}
	\label{fig:example}
\end{figure}

\begin{figure*}[!t]
	\centering
	\includegraphics[width=6.0in]{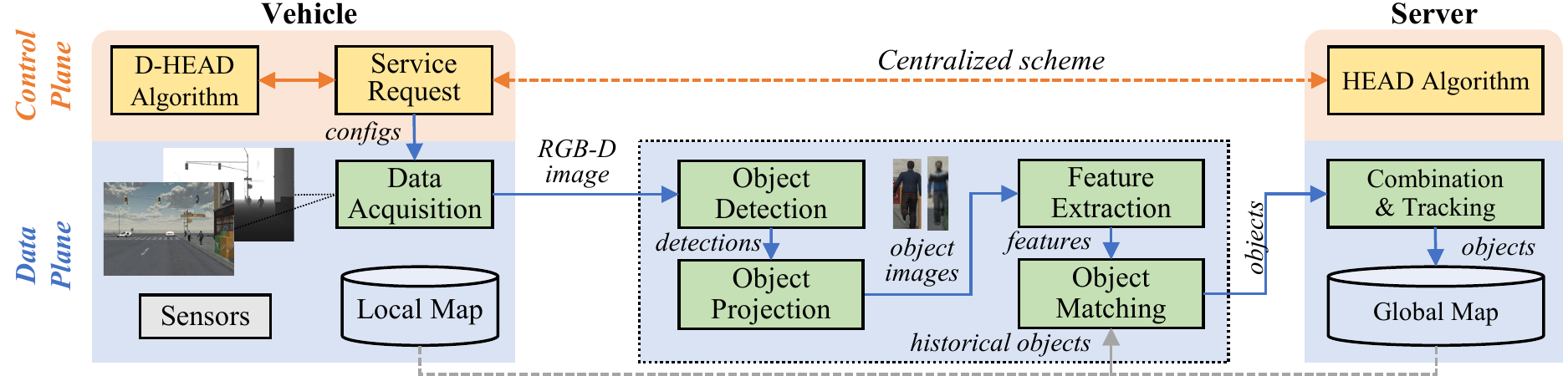}
	\vspace{-0.05in}
	\caption{\small The overview of \emph{LiveMap}. The data plane is to process sensor data for detecting, matching and tracking objects. The control plane is to manage networks for accelerating transmission and computation of vehicle offloadings.}
	\label{fig:architecture}
\end{figure*}

%%%%-----------  connected vehicle -------------%%%%
Connected and automated vehicles (CAVs) emerge in recent years to connect vehicles~\cite{guanetti2018control, rios2016survey} via advanced wireless technologies, e.g., 5G and beyond, with pervasive edge computing infrastructures~\cite{mao2017survey, shi2016edge}, e.g., edge servers in radio access networks (RAN).
The Automotive Edge Computing Consortium estimates that more than 50\% of all cars on the road in the United States will have connected features by 2025~\cite{AECC}.
Various onboard sensors of vehicles, e.g., cameras and LiDAR, can be leveraged to construct global information via crowdsourcing.
By using edge servers as the hub, the information perceived by individual vehicles is seamlessly collected, processed, and shared among vehicles and infrastructures with ultra-low latency.
% For example, if the car can obtain the perception of the truck, collision can be avoided 
% it can decide not to pass the truck even if the opposite lane is unobserved by the car.

% Connected vehicle is the key building block of the Internet of Vehicles (IoV)~\cite{contreras2017internet}, which connects vehicles with wireless technologies, e.g., cellular networks and dedicated short-range communications (DSRC).
% It allows the communication of vehicle-to-vehicle (V2V), vehicle-to-infrastructure (V2I), and vehicle-to-network (V2N), and could substantially improve driving safety by effectively sharing information among vehicles.

%%%%-----------  hard to share information -------------%%%%
However, it is non-trivial to share perception data among CAVs because of the constrained network infrastructures and resources (e.g., spectrum and servers).
For example, the perception of vehicles may have duplicated information due to their heavily overlapped sensing ranges in a dense urban scenario. 
In addition, the uplink transmission of vehicle perception data, e.g., point clouds, demands a tremendous data rate which may overwhelm mobile networks~\cite{ahmad2020carmap}.
Edge servers, that support hundreds of vehicles if not more, experience fast-changing traffic and workloads under varying vehicle trajectories.
Therefore, it is imperative to design intelligent network management solutions to achieve real time perception sharing under constrained network resources in automotive edge computing.

%%%%----------- livemap dynamic layer in edge server -------------%%%%
In this paper, we propose \emph{LiveMap}, a new real-time dynamic map as shown in Fig.~\ref{fig:example}. \emph{LiveMap} achieves the detection, matching, and tracking of objects on the road in the time scale of subseconds via crowdsourcing data from CAVs.
We design \emph{LiveMap} to achieve an efficient data plane for vehicle data processing and an intelligent control plane for vehicle offloading decisions.
The data plane is composed of object detection, object projection, feature extraction, object matching, and object combination.
In particular, we design to improve the object detection with new neural network pruning techniques, build concise feature extraction with variational autoencoder techniques, optimize the feature matching with a novel location-aware distance function, and increase the combination accuracy with a new confidence-weighted combination method.
The control plane enables adaptive vehicle offloading, e.g., offloading the computations from vehicles to servers, under varying network dynamics.
We design two algorithms, that apply in central and distributed scenarios, to minimize the latency of offloadings while satisfying the requirement of map coverage.
We design these two algorithms based on deep reinforcement learning (DRL) that optimize the vehicle scheduling and offloading decision of individual CAVs.
In addition, we implement \emph{LiveMap} on a small-scale physical testbed with multipleJetRacers (Nvidia Jetson Nano), a 5GHz WiFi router, and an edge server with Nvidia GPU.
% The performance of \emph{LiveMap} is comprehensively evaluated via both experiments in the testbed and simulations in the network simulator. 

The main contributions of this paper are listed:
\begin{itemize}[leftmargin=*]
    \item We design a new real time dynamic map (\emph{LiveMap}) via crowdsourcing sensor data of CAVs in automotive edge computing networks. 
    \item We develop an efficient data plane with sequential processing of sensor data for reducing the processing delay and improving detection accuracy.
    \item We design an intelligent control plane with two new algorithms that improve the latency performance without compromising the map coverage.
    \item We develop an edge network simulator and prototype \emph{LiveMap} on a small-scale physical testbed.
    \item We evaluate \emph{LiveMap} via both experiments and simulations, and the results validate its superior performance.
\end{itemize}

\vspace{-0.1in}
\section{\emph{LiveMap} Overview}
\label{sec:system_overview}
In Fig.~\ref{fig:architecture}, we overview the architecture of \emph{LiveMap}, which includes the data plane, i.e., sensor data processing for detecting, matching and tracking objects, and the control plane, i.e., network management for accelerating transmission and computation. 

The data plane is composed of several sequential processing components.
The acquisition component retrieves RGB-D images from CAV sensors and its relocalization.
The detection component detects possible objects in RGB images by exploiting state-of-the-art object detection framework, i.e., YOLOv3~\cite{redmon2018yolov3}.
The projection component projects the detected objects from pixel coordinates to world coordinates based on the depth information and camera-to-world transformation matrix.
The extraction component extracts visual features from cropped object images by using a variational autoencoder.
The matching component matches detected objects in either the local or global map according to their visual features and geo-locations.
The combination component combines multi-viewed objects by integrating a variety of attributes, e.g., confidence and geo-location.
Note that all components, except acquisition and combination, can be flexibly executed in either CAVs or edge servers, according to the control plane.
Finally, the global map will be updated, where new updates will be broadcasted to all vehicles for updating their local maps.

The control plane includes a central and a distributed scheme. 
In the central scheme, a vehicle sends a service request along with its local state to the edge server.
The HEAD algorithm optimizes the scheduling of this vehicle under the current map coverage, and determines the offloading decision with a central DRL agent under the global state if the vehicle is scheduled.
In the distributed scheme, the vehicle invokes the D-HEAD algorithm independently to optimize its scheduling and offloading decision according to the current local state.
The vehicle starts the data plane according to the offloading decision if it is scheduled.

\vspace{-0.1in}
\section{The Design of Data Plane }
\label{sec:data_plane}
The data plane is designed to efficiently process vehicle data in terms of processing delay and detection accuracy. 

\vspace{-0.1in}
\subsection{Data Acquisition}
We develop the acquisition component to acquire the sensor data, i.e., LiDAR and RGB-D images.
Without loss of generality, we consider the RGB and depth images from RGB-D cameras, e.g., Intel RealSense D435.
Besides, it obtains the accurate vehicle location in world coordinates, which depends on either high-accuracy GPS or advanced relocalization algorithms such as ORB SLAM2~\cite{murORB2}.
The accurate vehicle location is necessitated for combining the multi-viewed objects detected by multiple vehicles.

\vspace{-0.1in}
\subsection{Object Detection}
We design the detection component to detect transportation objects from RGB images, e.g., trucks and pedestrians, where detection results include the object classes, probability, and 2D bounding boxes.
Existing deep neural network (DNN) based algorithms, e.g., Fast-RCNN~\cite{girshick2015fast}, YOLO~\cite{redmon2018yolov3}, and SSD~\cite{liu2016ssd}, are mainly designed for generic classes with up to hundreds of object classes, e.g., book, kite, cup, and boat.
In the scenario of transportation systems, a large proportion of these generic classes would not appear on the road and are not interested in \emph{LiveMap}, e.g., books and fruits.
In general, the size of DNN increases in order to achieve similar detection accuracy under a larger number of classes, e.g., mean average precision (mAP).
As a result, the detection delay increases when these algorithms are directly applied to embedded platforms on vehicles.

To this end, we propose a slim detector with specified transportation classes to decrease the detection time while maintaining the detection accuracy.
We leverage the neural network pruning technique, which aims to decrease the size of DNN by pruning unnecessary neurons without compromising the detection accuracy noticeably.
We adopt the pruning workflow in~\cite{Liu_2017_ICCV} to iteratively prune the DNN with sparsity retrain, channel prune, and finetune.
First, the loss of the DNN retrain includes a weighted L1 regulation on the scaling factors in batch normalization (BN) layers and is
\begin{equation}
    Loss = \sum\nolimits_{(x,y)}{l(f(x|\mathbf{W}), y)} + \lambda \sum\nolimits_{\gamma} g(\gamma),
\end{equation}
where $\mathbf{W}$ is the DNN weights, $\lambda$ is a balancing factor, and $x, y$ are the input images and ground-truth labels, respectively.
The function $l(\cdot)$ is the original loss function, and $g(\gamma)$ is the sparsity-induced penalty on the scaling factors, where $\gamma$ is the scaling factor of the channel in convolutional layers.
Second, as the sparsity training completes, these insignificant convolutional channels (nearly zeros scaling factors) are removed in the channel pruning.
Third, the DNN is further fine-tuned to re-gain the accuracy performance. 
Note that these processes repeat to trade-off between accuracy and size of the DNN.

The performances of our object detector (based on YOLOv3 tiny~\cite{redmon2018yolov3}) are as follows.
The detection classes are reduced from 80 to 10, and the network size is reduced from 8.69M to 0.54M (93.7\%) under 0.01 mAP degradation (from 0.534 to 0.524) on our dataset. 

\begin{figure}[!t]
	\centering
	\includegraphics[width=2.8in]{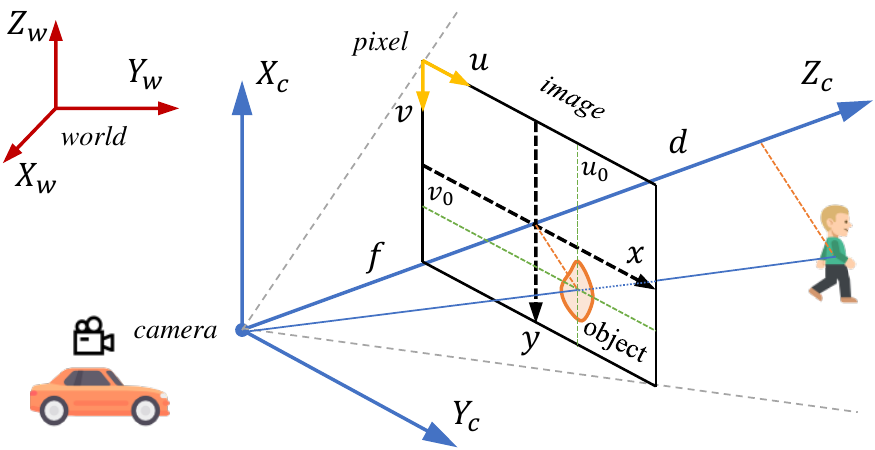}
	\vspace{-0.05in} \caption{\small The camera coordinate system.}
	\label{fig:projection}
\end{figure}

\vspace{-0.1in}
\subsection{Object Projection}
We design the projection component to derive the object location in the world coordinate system according to the detection results and depth information.
In Fig.~\ref{fig:projection}, we illustrate the projection of objects in camera coordinate systems.
Thus, we can calculate the object location in world coordinates by transforming pixel-to-camera and camera-to-world accordingly.

First, we transform the locations from pixel to camera coordinates, given the camera focal length $f$ and the image resolution $(R_W, R_H)$.
Denote $(u_0, v_0)$ as the central pixel location of an object, the 3D location\footnote{The conversion from pixel-to-camera coordinates may vary according to the axis setting of the camera coordinate system.} $(X, Y, Z)$ in the camera coordinates are
\begin{align}
    X \; &= \; - (d*(v_0 - 0.5*R_H)) / f, \nonumber \\ 
    Y \; &= \; (d*(u_0 - 0.5*R_W)) / f, \\
    Z \; &= \; d, \nonumber
\end{align}
where $d$ is the object depth in the Z-axis.
However, calculating the object depth is non-trivial, because objects usually occupy irregular areas while its bounding box is a rectangle.
In traditional depth calculation approaches, all the depth values in the bounding box of the object are averaged.
These approaches would lead to an inaccurate estimation of the object depth, because the object usually does not cover all the pixels in the rectangle bounding box.
To address this issue, we randomly sample multiple small squares (e.g., 4x4) around the center of the bounding box.
The average depth of all these small squares are sorted, where the maximum and minimum value are removed.
The final depth is then the average depth of these remaining small squares.
This method achieves a more robust estimation of object depth and improves the accuracy of detected objects eventually.

Second, we transfer the location in camera coordinates to world coordinates. The world location \textbf{$(W_x, W_y, W_z)$ } are
\begin{equation}
    \left[W_x,\; W_y,\; W_z,\;1\right]^T = M_{c2w} \times \left[X,\;Y,\;Z,\;1 \right]^T,
\end{equation}
where $M_{c2w}$ is the conversion matrix.

\vspace{-0.1in}
\subsection{Feature Extraction}
We design the extraction component to extract concise visual features from the cropped object images by using the technique of variational autoencoder.
Even if we obtain the object location in world coordinates through the aforementioned components, we still do not know who are them and where did they from.
We need further identify them for tracking their trajectories consistently, which necessitates the feature extraction.
Traditional feature extraction methods (e.g., ORB~\cite{rublee2011orb}) cannot apply here for two reasons.
First, they may generate a similar amount of features as compared to that of object images~\cite{zhang2018jaguar}, which worsens not only the following matching time but also the transmission delay of possible vehicle offloading.
Second, they generate very limited features for small object images, which substantially compromises the accuracy of the following feature matching.
In the transportation scenario, the dimension of the cropped objects are typically small as they are far away from vehicles, e.g., the image size of a truck may be 50x100 out of 1080p images.

% , since their loss functions are calculated only based on the reconstruction error between the input and rebuilt image, 
% This irregular latent space breaks the correlations among images in terms of Euclidean distance.

To this end, we exploit the variational autoencoder~\cite{kingma2013auto} to derive condensed features from small images.
The autoencoder is composed of an encoder (encoding input image into features) and a decoder (rebuilding the image from features).
The variational autoencoder overcomes the deficiencies of traditional autoencoders, i.e., irregulate latent space~\cite{kingma2013auto}, by using a new regularized loss function.
Given the image $x$ and sampled latents $z$ from the distribution $\mathcal{N}(\mu, \sigma^2)$, the loss function of the variational autoencoder is expressed as 
\begin{equation}
    Loss = - \mathbb{E}_{z \sim q(z|x)} \left[  \log p(x|z)  \right] + D_{KL} \left[ q(z|x) | p(z)\right], 
\end{equation}
where $q(z|x)$ and $p(x|z)$ are the encoder and decoder, respectively. The $D_{KL}$ is the KL-divergence, which evaluates the discrepancy between two distributions, and $p(z) \sim \mathcal{N}(0,1)$ is a Gaussian distribution.
We collect all the object images detected by vehicles, and train the variational autoencoder accordingly.
During the inference phase, the generated features will be treated as the features of the object.

\vspace{-0.1in}
\subsection{Object Matching}
We design the matching component to compare and match detected objects in the map according to generated features and geo-locations.
Because a map may include hundreds or thousands of objects, matching an object with all these objects in the map is time-consuming and inefficient.
Besides, recent researches~\cite{ahmad2020carmap} indicate that the feature distance (i.e., features generated by the feature extraction component) may fail in transportation scenarios.

To this end, we use a new location-aware function to measure the distance between two features and improve the robustness of feature matching in three steps.
First, we select the candidate matching objects whose geo-distance is less than 100 meter with the detected object, which helps to reduce the size of candidate objects for matching.
Second, we build a trajectory model for each object in the map, which is fitted based on the historical locations and used to predict the future location of the object.
Third, we develop the following distance function to include both the feature distance and geo-distance of the $i$th and $j$th object, which is
\begin{equation}
    D_{i,j} = \min(\left[||z_{i,m}-z_{j,m} ||^2, \forall m \in \mathcal{M}\right]) + w ||g_i -g_j||^2, \label{eq:matching}
\end{equation}
where $g$ is the location, $w$ is a weight factor, $z$ are the object features, $||\cdot||^2$ is the L2-norm operation, and $\mathcal{M}$ denotes the set of features of an object.
Note that an object could be viewed by different vehicles, these multi-view features are considered as valid and associated with the object.

\vspace{-0.05in}
\subsection{Object Combination}
We develop the combination component to integrate and update the detected objects viewed by different vehicles in the global map.
The global map includes all the active objects, where each object includes multiple attributes such as \emph{object id}, \emph{object name}, \emph{geo-location}, and \emph{feature}.
Note that the objects with outdated locations are automatically removed from the global map. 

Note that the multi-viewed objects may have different results, e.g., locations, confidences, and features.
Thus, we use a new combination method to calculate the geo-location of objects as
\begin{equation}
    g =  \sum\nolimits_{m \in \mathcal{M}} { (\mathcal{P}_m * g_m) }/{(\sum\nolimits_{m \in \mathcal{M}} \mathcal{P}_m )} ,
\end{equation}
where $\mathcal{P}_m$ and $g_m$ are the detection confidence and geo-location, respectively.
In addition, the multi-view features of the object are considered to be valid, which are included in the map for achieving better matching accuracy as shown in Eq.~\ref{eq:matching}.
Finally, only the newly updated information in the global map, e.g., pedestrian features, are broadcasted to all the CAVs.

%  with some vehicle information, e.g., wireless qualities and computation capacities
\vspace{-0.1in}
\section{The Design of Control Plane }
The control plane is designed to improve the system performance in terms of latency and coverage.
We introduce the system model, formulate the problem, and propose new algorithms in both central and distributed scenarios.
\vspace{-0.1in}
\subsection{System Model}
We consider multiple CAVs that connect to a cellular base station and an edge server.
To complete the dynamic map, all vehicles asynchronously offload their computation tasks, i.e., the data plane, to the edge server.
As these tasks are naturally separated, we consider the discrete partition between these components.
We denote $\mathcal{I}$ as the set of vehicles and $\mathcal{N} = \{0, 1, ..., N\}$ as all the possible partitions, where $N$ is the maximum partition.
Specifically, we define $y_i$ as the partition of the $i$th vehicle, i.e., offloading decision.
For instance, the partition is $2$ suggests that the detection and projection component in \emph{LiveMap} are completed on the vehicle.
Then, the generated intermediate data, e.g., cropped object images, are transmitted to the edge server for further computation.
Next, the remaining extraction, matching and combination components are executed on the edge server.

We consider individual vehicles have a geographic coverage area, e.g., a circle with a 50m radius, depending on the sensing range of onboard sensors and the vehicle locations. 
Denote the coverage of the $i$th vehicle at time slot $t$ is denoted as $C_i^{(t)}$.
To avoid excessive computation offloading from vehicles with heavy coverage overlap, we introduce the vehicle scheduling indicator to determine if the vehicle is allowed to offload. 
We define the scheduling indicator of the $i$th vehicle as $x_i \in \{0, 1 \}$, where $x_i=1$ suggests that the vehicle is scheduled to conduct the offloading.
For the sake of simplicity, we denote $\mathcal{X}$ as the set of vehicle scheduling and $\mathcal{Y}$ as the set of offloading decision, respectively.
The latency $L_i^{(t)}$ of the $i$th vehicle at time slot $t$ is defined as the elapsed time since the vehicle starting its offloading until the map updates are broadcasted.

\vspace{-0.1in}
\subsection{Problem Formulation}
The objective of the control plane is to provide real time environmental information to all the CAVs.
In the transportation scenarios, outdated transient information is less helpful for the decision of autonomous driving and ADAS.
For instance, the location of vehicles 30 seconds ago probably fail to contribute to the current control, e.g., lane changing.
Moreover, we aim to maintain the dynamic map with a large geographic coverage to provide comprehensive information to all the CAVs.
Hence, we denote the achievable instant map coverage as $\textstyle \bigcup\nolimits_{i\in \mathcal{I}} C_i^{(t)}$ at the time $t$, i.e., the union of coverage of all vehicles.
Then, we formulate the optimization problem of the control plane as
\begin{align}
\centering
\label{prob0}
&{\min \limits_{ \{\mathcal{X,Y}\}} \;\;\;\;\;\;  \sum\nolimits_{t \in \mathcal{T}}{\sum\nolimits_{i \in\mathcal{I}}L_i^{(t)}}} \\ 
&{\;\;s.t.\;\;\;\;\;\;\textstyle \bigcup\nolimits_{i\in \mathcal{I}, x_i \neq 0} C_i^{(t)} \ge \beta \textstyle \bigcup\nolimits_{i\in \mathcal{I}} C_i^{(t)}, \forall t \in \mathcal{T}}, \label{prob:const1} \\ 
&{\;\;\;\;\;\;\;\;\;\;\;\;\;x_i^{(t)} \in \{0, 1\}, \forall i \in \mathcal{I}},  t \in \mathcal{T}, \label{prob:const2} \\ 
&{\;\;\;\;\;\;\;\;\;\;\;\;\;y_i^{(t)} \in \{0, 1, ..., N\}, \forall i \in \mathcal{I}, t \in \mathcal{T}}, \label{prob:const3}
\end{align}
where we introduce $\beta \in [0, 1]$ as a factor to constrain the overall instant map coverage.
The $\mathcal{T}$ is a given time period such as 15 minutes or 1 hour to evaluate the statistical performance.

However, we observe that the above problem is challenging to be resolved.
First, the accurate mathematical model to describe the latency of a vehicle can hardly be obtained in real network systems, i.e., $L_i^{(t)},\forall i \in \mathcal{I}, t \in \mathcal{T}$ are unknown.
On the one hand, the asynchronous mechanism of vehicle offloadings inevitably leads to overlap among offloadings and resource competition in the time domain.
The shared radio transmission, e.g., cellular networks, and server computation, e.g., edge computing, are not considered to be virtualized and controlled, from the perspective of operating \emph{LiveMap}.
On the other hand, vehicles are heterogeneous in terms of the capability of onboard hardware, routes, and the quality of wireless channels.
As a result, the resource competition is fast-changing, whose status, e.g., which vehicles are uplink transmitting and how they share radio resources, are unable to be observed.
Second, the offloading of all vehicles exhibit the \emph{Markov} property.
For example, the decision made for a vehicle at the current time not only affects the current performance of vehicles but also influences the further system states.

\vspace{-0.05in}
\subsection{Centralized Algorithm Overview}
In this part, we overview the centralized HEAD algorithm based on deep reinforcement learning (DRL) to effectively solve the above problem in our previous work~\cite{liu2021livemap}.
In the HEAD algorithm, we solve this problem by alternatively tackling the vehicle scheduling and offloading decision.
We observe that the offloading of vehicles are usually completed in subseconds, in contrast, the scheduling of vehicles may run at the higher time scales, e.g., seconds.
Therefore, we design a hierarchical algorithm named HEAD as follows.
In the outer layer, we determine the scheduling of vehicles by minimizing the total number of scheduled vehicles while maintaining the constraint of overall map coverage.
Note that excluding more vehicles suggests that there will be less competition for network resources among vehicles, which thus accelerates the transmission and computation of vehicle offloadings in general.
In the inner layer, we leverage the DRL technique to intelligently optimize the offloading decision for individual scheduled vehicles, where the dimension of the state and action turns out to be fixed.

\begin{algorithm}[!t]
	\caption{The D-HEAD Algorithm}\label{alg:proposed-dist}

	\KwIn{ $\beta$, $\theta^*, i$,}
	\KwOut{$x_i, y_i$}

    	$\mathbf{s}_{t} \gets [\mathbf{s}^v_{t},\; y_{t-1},\;  L^{(t-1)}]$, $/*\;build\; state\; */$\;
    	$C_k \gets C_i,\; \forall k \in \mathcal{I}$, $/*\;estimate\; coverage\; */$\;
    	\If {time to schedule}
    	{
		    $x_k \gets 1,\; \forall k \in \mathcal{I}$\;
        	\While{$True$}
        	{
        	 $k \gets \arg\max\nolimits_{k \in \mathcal{I}, x_k \neq 0}O_k$\;
        	 $x_k \gets 0$;
        	 
        	 \If {$\bigcup\nolimits_{k\in \mathcal{I}, x_k \neq 0} C_k^{(t)} \leq \beta \bigcup\nolimits_{k\in \mathcal{I}} C_k^{(t)}$}
        	    {
        	        $x_k \gets 1$\;
        	        \textbf{break}\;
        	    }
        	}
    	}
    	\If {$x_i == 1$ (scheduled)}
    	{
        	$y_i$ $\gets$ $\arg\max \nolimits_{\mathbf{a}_{t}}Q^*(\mathbf{s}_{t}, \mathbf{a}_{t} | \theta^*)$, $/*\;get\; action */$\;
    	}
    	\Else
    	{
    	    $y_i \gets -1$,  $/*\;not\; scheduled*/$\;
    	}
    % 	vehicle $\gets$ $(x, y)$, $/*\;return\; to\; vehicle */$\;
	
    \Return{$x_i, y_i$}\;
\end{algorithm}

\vspace{-0.05in}
\subsection{Distributed Algorithm Design}
In this part, we design a new distributed HEAD (D-HEAD) algorithm (Alg.~\ref{alg:proposed-dist}) to solve the problem, which alleviates the need for central decisions in the aforementioned centralized algorithm.
The idea of the distributed algorithm is to allow individual vehicles to optimize their scheduling and offloading according to their local states, e.g., local map.
The distributed algorithm follows the two-layer framework as the centralized algorithm does, i.e., the vehicle scheduling and offloading decision occur in the upper-layer and lower-layer, respectively.
In particular, individual vehicles optimize their vehicle scheduling to determine if they participate in the offloading. 
This vehicle scheduling is designed to be asynchronous, for avoiding the reduction of coverage requirements during the periodical scheduling interval due to the mobility of vehicles (see Fig.~\ref{fig:simulation_coverage_time}).
The offloading decisions are optimized (if scheduled) locally whenever there is an offloading to start.
In this way, the communication overhead (incurred by transmitting states to the central controller) and action delays (the delay to send the state and receive the action from the central controller) are eliminated.
However, the lack of timely global states may compromise the overall performance of the algorithm.

\subsubsection{Vehicle Scheduling}
The challenge of individual vehicle scheduling is the unknown coverage and scheduling decisions of other vehicles, even if the historical vehicle locations may be obtained from its local map.
For example, if nearby vehicles are scheduled, this vehicle with heavy coverage overlap may be exempt from participating offloadings and thus alleviate the complex competition of network resources.
We propose a distributed scheduling method in three steps to address this challenge.

First, we predict the location of all other vehicles at the next time slots based on the local map in individual vehicles.
As the local map stores the historical trajectory of every detected vehicles, we adopt the simple but effective linear prediction to predict the location of vehicles as follows
\begin{equation}
    g(t+1) = 2\cdot g(t) - g(t-1).
\end{equation}
where $g(t)$ is the 2-dim world location of a vehicle.
Although we adopt this linear prediction in this work, other advanced trajectory prediction methods, e.g., LTSM, can be further applied to improve the accuracy of prediction. 

Second, we construct a complete graph $(V, E)$ to represent the correlation of coverage among vehicles.
We denote the vertices $V$ as all the vehicles and the edges $E$ as the overlapping of vehicle coverages.
In particular, the edge value between $i$th and $j$th vertex (i.e., vehicle) are calculated as 
\begin{equation}
    e_{i,j} =  {(C_i \bigcap C_j)}/{(C_i \bigcup C_j)},
\end{equation}
where $e_{i,j}=e_{j,i}$. 
As the vehicle cannot obtain the accurate coverage range of all other vehicles, we assume all vehicles have the same coverage range as this vehicle.
We will evaluate the effect of heterogeneous coverage range on the effectiveness of vehicle scheduling (see Fig.~\ref{fig:simulation_coverage_range}).
Then, we define the average overlapping ratio (AoR) of the $i$th vehicle as 
\begin{equation}
    O_i = {1}/{|\mathcal{I}|}\sum\nolimits_{j \in \mathcal{I}, j \neq i} e_{i,j}.
\end{equation}

Third, we iteratively prune the vertex in the graph $(V, E)$ with the highest AoR, i.e., $i = \arg\max\nolimits_{k \in \mathcal{I}}O_k$.
The basic idea is that we gradually exclude a vehicle from participating offloading at the minimum cost of the overall map coverage.
With all the generated intermediate graphs, we search for the optimal graph which achieves the minimum number of vertices and meets the constraint of overall map coverage.

Note that all individual vehicles follow this method to determine their scheduling decisions independently.
If this vehicle is not in the generated list of scheduled vehicles, then it will continue to process all the computation components locally, i.e., its offloading decision is the maximum by default.
This vehicle scheduling process is initialized when all the computations of this vehicle are completed in either the local vehicle or the remote edge server. 

% vehicle moves

\subsubsection{Offloading Decision}

The vehicle offloading is complicated in dynamic automotive edge computing networks, which depends on not only the high-dim network states but also the complex and unknown resource competitions among vehicles.
Therefore, we propose to exploit deep reinforcement learning~\cite{schaul2015prioritized} to determine the offloading decision for individual vehicles.
As the aforementioned vehicle scheduling assures the constraint of map coverage, the optimization of offloading decision becomes an unconstrained problem with a fixed size of state and action space.

As the offloading decision is made in individual vehicles, the offloading problem falls in the multi-agent reinforcement learning (MARL) setting.
There are multiple agents $\mathcal{N}$ (each in scheduled vehicles) that interact with the environment asynchronously.
At every decision time $t$, individual agent can observe the network state $\mathbf{s}_{t}$ and needs to take an offloading action $\mathbf{a}_t$.
Then, the agent will receive a reward $r(\mathbf{s}_t, \mathbf{a}_t)$, and the environment transits to the next state $\mathbf{s}_{t+1}$ accordingly.
The objective is to find policies $\pi^*=\{\pi_n^*, \forall n \in \mathcal{N}\}$ for all agents to map local states to actions and maximizes the discounted cumulative reward $R_0 = \sum\nolimits_{t=0}^{\infty} \gamma^t r(\mathbf{s}_t, \mathbf{a}_t)$. Here, $\gamma \in [0, 1)$ is a non-negative discounting factor.

The challenge of solving the MARL lies in the interrelations among these agents and the lack of global states in individual agents, e.g., the queuing vehicles on the edge server.
We design a distributed method by following the centralized training and distributed execution framework to solve the MARL problem.
In particular, a common policy is created in the central edge server during the training phase, which is updated by gathering all the transitions from individual agents in vehicles.
Then, the updates of the common policy are sent back to individual agents.
As the transitions are available in the edge server, the common policy is trained by following the procedures in the central algorithm, i.e., DQN~\cite{mnih2015human} with prioritized experience replay (PER)~\cite{schaul2015prioritized}.
During the execution phase, all agents in vehicles share the same policy and optimize their offloading decision independently.

\textbf{State space} is designed to completely and concisely represent the status regarding vehicle offloadings. 
Specifically, the state space is defined as $\mathbf{s}_{t} \triangleq [\mathbf{s}_{t}^v,\; \mathbf{s}_{t}^s]$.
The $\mathbf{s}_{t}^v$ is the status of the vehicle, which includes its quality of wireless channel (e.g., RSSI) and its capability of hardware (e.g., CPU architecture, CPU cores and frequency, GPU architecture and frequency).
The $\mathbf{s}_{t}^s$ is the status of the system, which incorporates the capability of the edge server, the number of total connected and queued vehicles and the total wireless bandwidth of the base station.

\textbf{Action space} is designed to enable adaptive vehicle offloading, i.e., different partitions of computation tasks of vehicles, which is $\mathbf{a}_t \triangleq [ y ]$.

\textbf{Reward} is designed to guide the training of the DRL agent for reducing the cumulative latency of vehicle offloadings, which is the negative offloading latency, i.e., $r(\mathbf{s}_t, \mathbf{a}_t) \triangleq - L^{(t)}$.

\vspace{-0.1in}
\section{System Implementation}
\label{sec:implementation}
We develop the system prototype of \emph{LiveMap} on a small-scale physical testbed and a network simulator for large-scale evaluations.

\vspace{-0.1in}
\subsection{System Prototype}
We build a small-scale testbed to implement \emph{LiveMap} and evaluate the efficiency of the data plane.
We use four JetRacers as the CAVs, where each JetRacer has an embedded Nvidia Jetson Nano.
We use a 5GHz WiFi router to connect the edge server with Intel i7 and Nvidia GTX 1070~\cite{nvidia2011nvidia}.
To emulate wireless channel dynamics, we dynamically change the transmit power (from 1dBm to 22dBm) of both the transmitter and receivers with Linux "\emph{iw}" command.
In addition, we develop a FIFO queue on the edge server to serve all the incoming vehicle offloadings.
In addition, we compress the intermediate offloading data with the LZ4 compression.

% \begin{figure}[!t]
% 	\centering
% 	\includegraphics[width=3.4in]{./fig/testbed.pdf}
% 	\vspace{-0.05in} \caption{\small The implementation of \emph{LiveMap}.}
% 	\label{fig:unity_env}
% \end{figure}

% Autoencoder architecture
% To train autoencoder, we collect more than 10k object images from 
% As a result, we resize all object images to 50x50 resolution, feed them into the encoder network that generate latent vectors with 25 elements. The encoder network is designed with 3-layer convolution neural networks, i.e., (3, 16, 1), (16,64, 2), and (64, 256, 2).  

We develop DRL agents with Python 3.7 and PyTorch 1.4.
To train DRL agents, we adopt Deep Q network (DQN)~\cite{mnih2015human} with prioritized experience replay~\cite{schaul2015prioritized}.
In particular, we create the [256, 256] fully-connected DNN with the activation function of Leaky Recifier~\cite{goodfellow2016deep}.
We set the learning rate as 0.5e-3 with batch size 512 and $\gamma=0.9$.
Besides, a $\epsilon$-greedy exploration is applied, which will be decayed during the training.
% which starts from probability 0.5 to 0.1 during the training phase.

\vspace{-0.05in}
\subsection{Traces DataSet}
Both experiments and simulations require the vehicle data and traffic traces.
Thus, we build an environment to generate traces with Unity3d (i.e., modern city package) with a variety of transportation scenarios such as intersection and highway.
In each scenario, we simulate hundreds of pedestrians and vehicles with different velocities, where the paths of pedestrians and vehicles are predefined.
We collect more than 1K time stamps, where each time stamp includes the RGB-D image, ground-truth location, and camera-to-world transformation matrix of all vehicles.
Besides, the location of pedestrians are also recorded for evaluating the accuracy of \emph{LiveMap}.
Individual vehicles are with a front RGB-D camera (50mm focal length, $54.04^{\degree}$ FoV, 50m perception range), and the generated images are with 741x540 dimension.

\vspace{-0.05in}
\subsection{Network Simulator}
We build a time-driven simulator to simulate automotive edge computing networks.
The simulator includes vehicular computation modules, a radio transmission component, and an edge computation module.
We develop the radio transmission module based on a 5G system-level simulator~\cite{oughton2019open}, where the urban micro (UMi - Street Canyon) channel model~\cite{esti_tr_138_901} is adopted.
The total uplink and downlink bandwidth are equally shared by all the available vehicles (i.e., they are scheduled and currently transmitting data) for the sake of simplicity.
Both the onboard and server computation are simulated with FIFO queues.

The feature of this simulator lies in the \emph{tasks} object, where a task corresponds to the computation of a vehicle.
The task includes multiple variables, including the vehicular computation time, the uplink transmission data size, and edge computation time.
These data are obtained by sampling from real-world measurements in the testbed (see Fig.~\ref{fig:system_comparison}).
First, a task will be created once the offloading decision is determined, which starts from the onboard computation module of individual vehicles.
The task computation onboard is simulated by reducing the \emph{remaining\_onboard\_computation\_time} for every simulation time step such as 1 ms.
Then, the task is transferred to the next wireless transmission module if its \emph{remaining\_onboard\_computation\_time} is zero.
Accordingly, the task will be sent to the server computation if its \emph{uplink\_transmission\_data\_size} is reduced to zero.
The server computation is similar to that of onboard, but it has multiple parallel queues with a \emph{min\_load} queue scheduling method for assigning the incoming tasks to these queues. 
Eventually, the task is completed as its \emph{downlink\_transmission\_data\_size} becomes zero, and its latency will be recorded.

\vspace{-0.05in}
\subsection{Comparison Algorithms}
We compare \emph{LiveMap} with the following algorithms that schedule all vehicles:
\begin{itemize}[leftmargin=*]
    \item Edge offloading (\textbf{EO}): The offloading decision of vehicles in EO are $\mathbf{a}_t = 0, \forall t$, where all the computations are executed on the edge server.
    \item Local process (\textbf{LP}): The offloading decision of vehicles in LP are $\mathbf{a}_t = 4, \forall t$, where all the computations are completed on the vehicles.
    \item Random offloading (\textbf{RO}): The offloading decision of vehicles in RO are randomly selected.    
    \item Regression model (\textbf{RM}): RM relies on a multivariate polynomial regression model to predict the offloading latency under different network states. RM selects the offloading decision with the predicted minimum latency. Two factors are identified as the key network states, i.e., the number of CAVs and wireless conditions. The input dimension of RM includes the network states and the offloading decision. The RM model is trained with \emph{scikit-learn} tool, where the training dataset is collected from experiments.
    \item \textbf{LiveMap-Lite}: LiveMap-Lite optimizes offloading decision as \emph{LiveMap} does, but it schedules all vehicles. 
    \item \textbf{LiveMap-Dist}: LiveMap-Dist uses the D-HEAD algorithm to optimize the vehicle scheduling and offloading decisions.
\end{itemize}

\vspace{-0.1in}
\section{Performance Evaluation}
\label{sec:evaluation}
We evaluate the performance of \emph{LiveMap} through both experiments in the small-scale testbed and simulations in the network simulator.
The objective of the evaluation includes: 1) compare \emph{LiveMap} with state-of-the-art solutions in terms of overall performance; 2) quantify latency and coverage performance of the D-HEAD algorithm in the control plane under varying network dynamics; 3) justify the latency and accuracy performance achieved by the data plane. 
In the experiments, the constraint of overall map coverage $\beta=0.8$ and available offloading decisions are $\{0,1,2,3,4\}$.

\begin{figure}[!t]
	\centering
	\includegraphics[width=3in, height=1.8in]{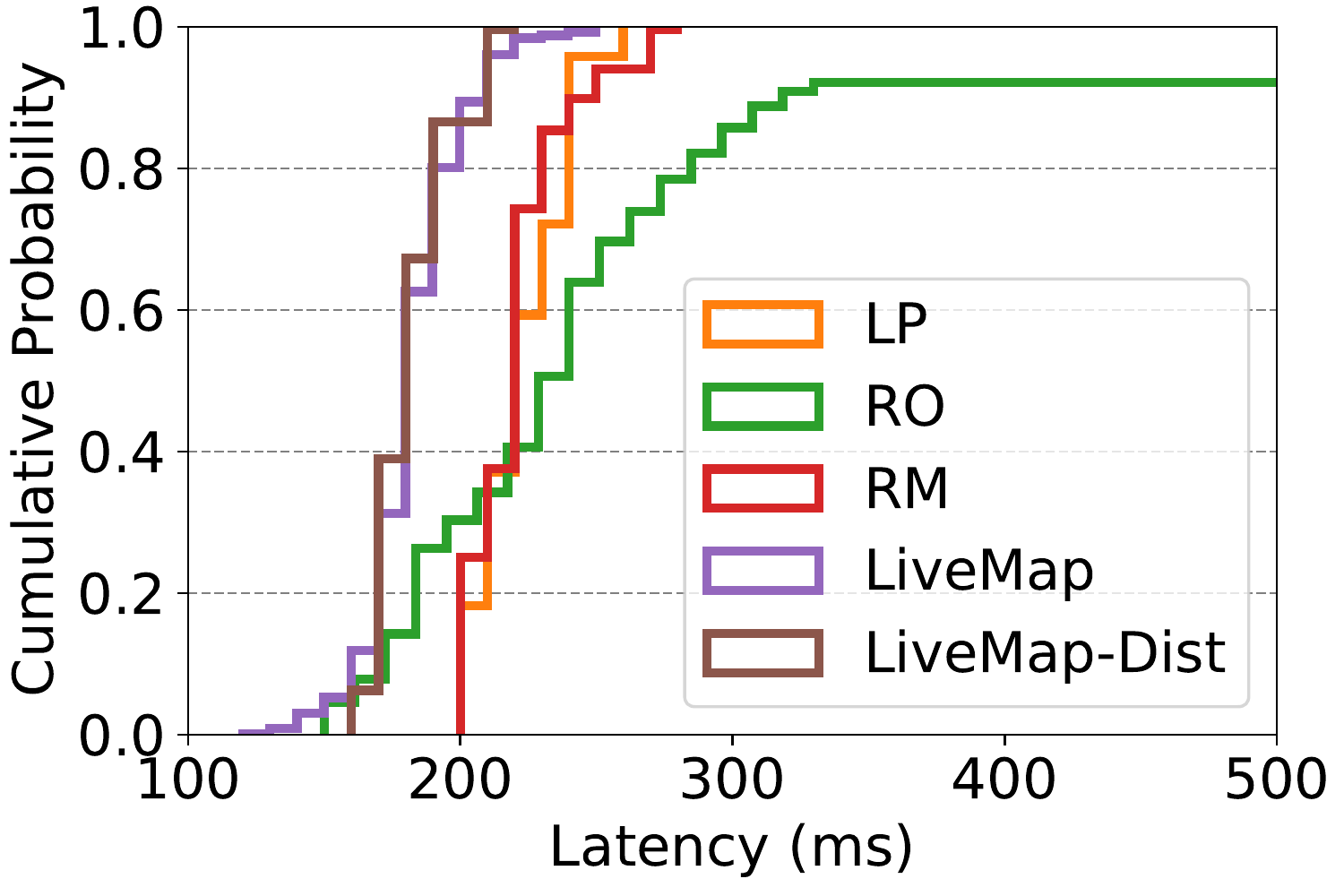}
	\caption{\small The cumulative probability of latency.}
	\label{fig:latency_cdf}
\end{figure}

\begin{figure*}[!t]
    \begin{minipage}[!t]{0.32\linewidth}
        \centering
    	\includegraphics[width=2.3in]{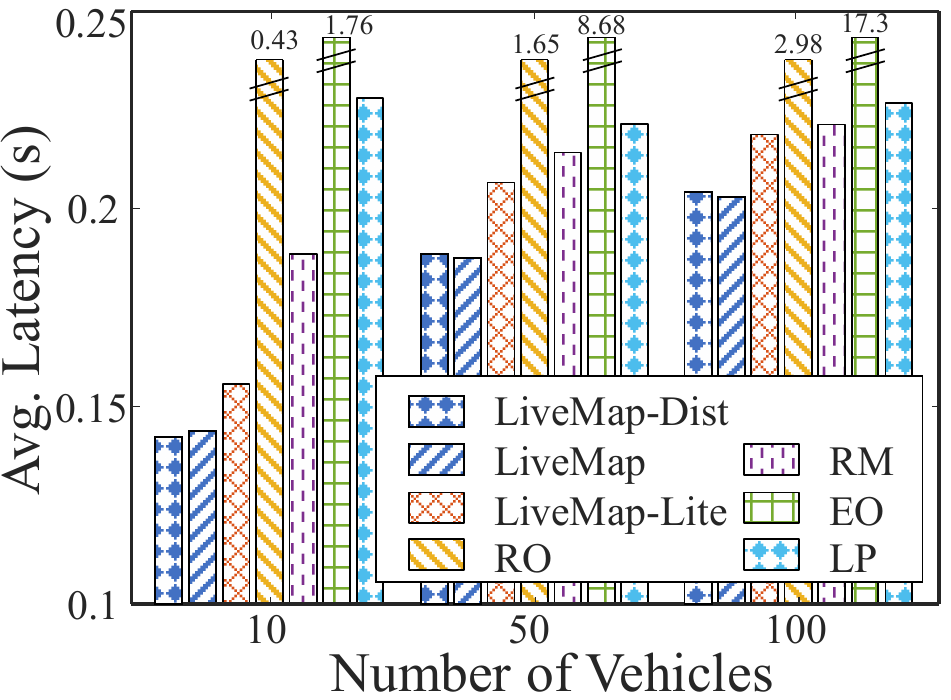}
    	\caption{\small Latency under different number of CAVs.}
    	\label{fig:simulation_carnum_coverage_1}
    \end{minipage}
\hfill
    \begin{minipage}[!t]{0.32\linewidth}
        \centering
    	\includegraphics[width=2.3in]{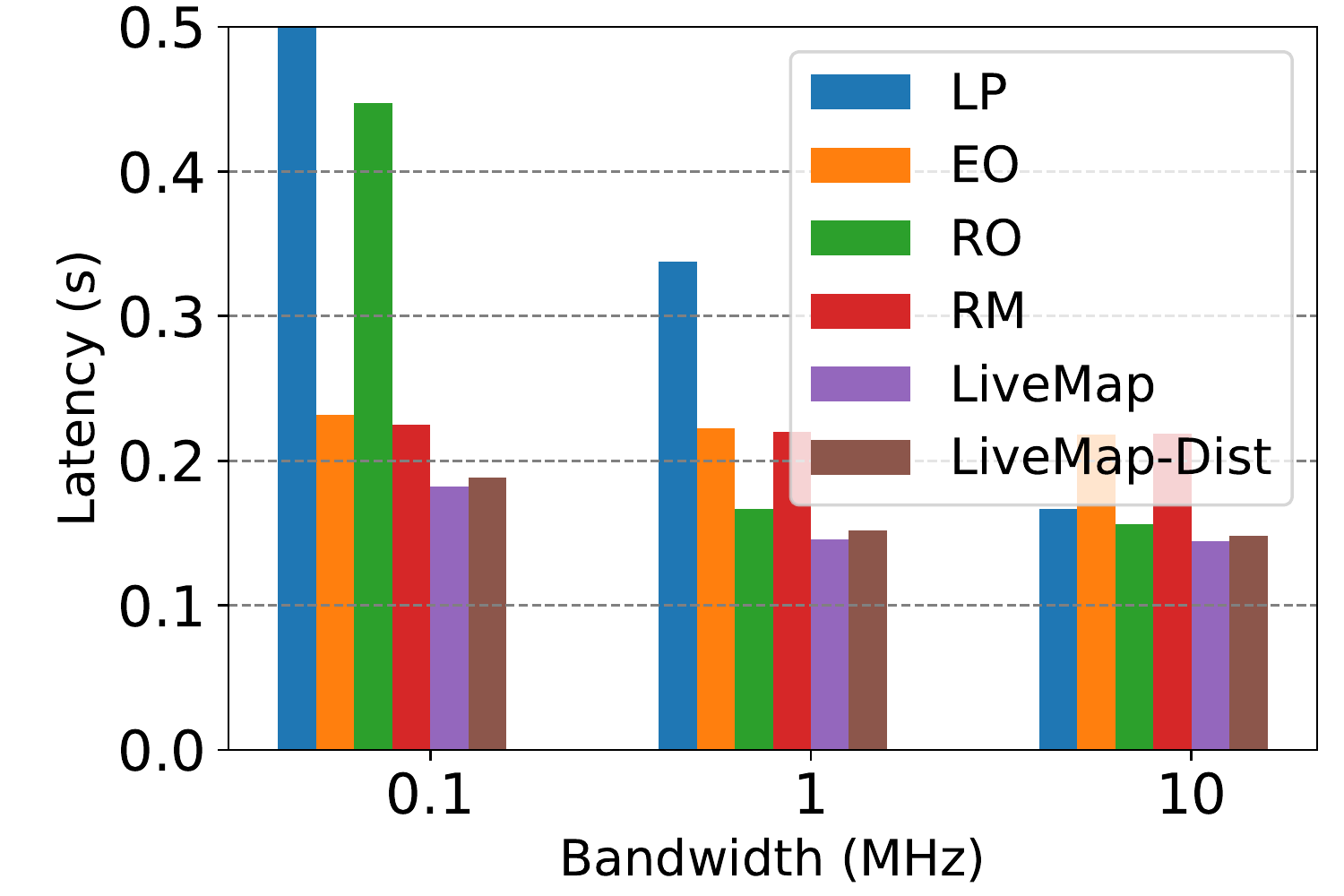}
    	\caption{\small Latency under different wireless bandwidth.}
    	\label{fig:simulation_bandwidth}
    \end{minipage}
\hfill
    \begin{minipage}[!t]{0.32\linewidth}
        \centering
    	\includegraphics[width=2.3in]{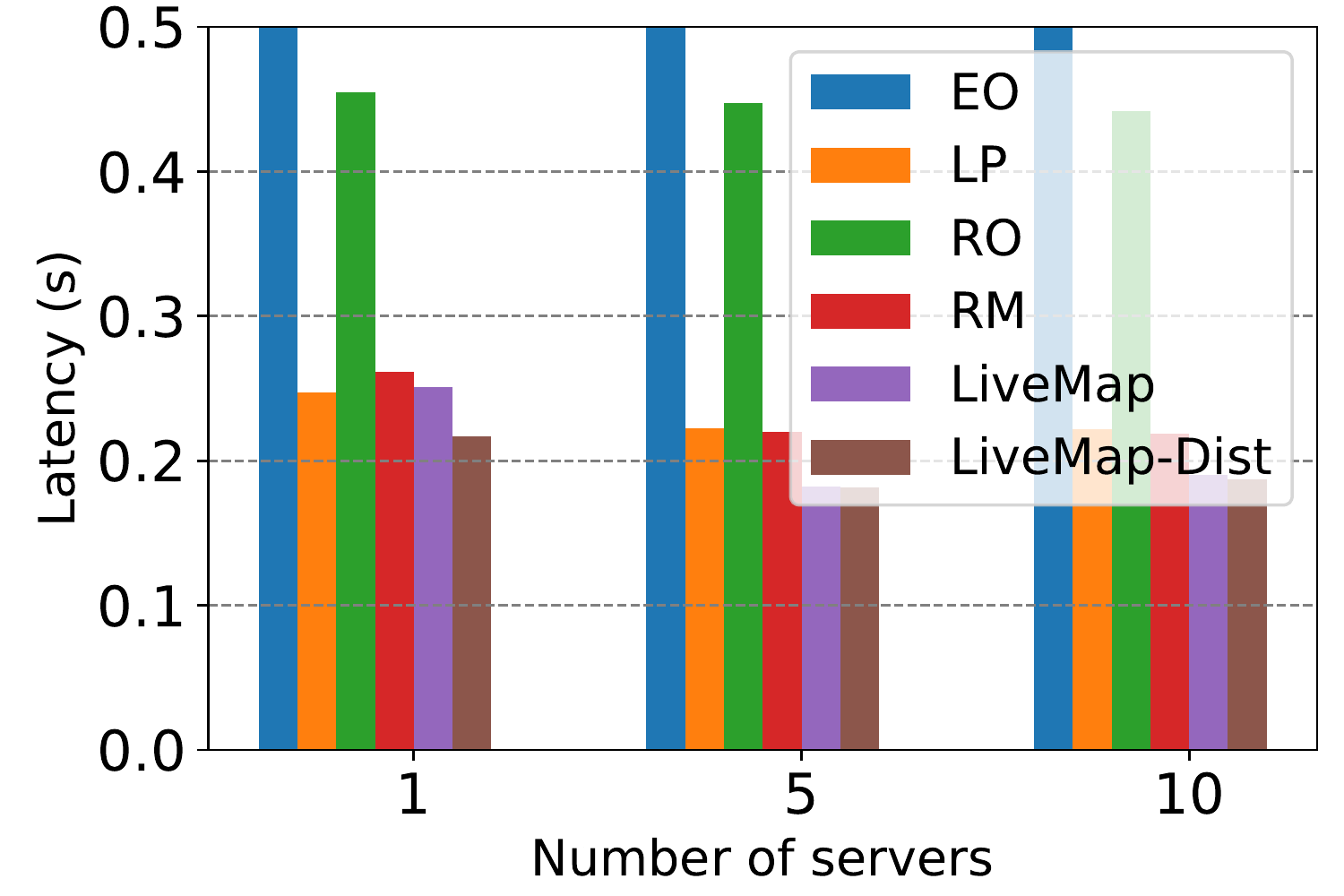}
    	\caption{\small Latency under different number of servers.}
    	\label{fig:simulation_num_server}
    \end{minipage}
\end{figure*}

\begin{figure*}[!t]
    \begin{minipage}[!t]{0.32\linewidth}
        \centering
    	\includegraphics[width=2.3in]{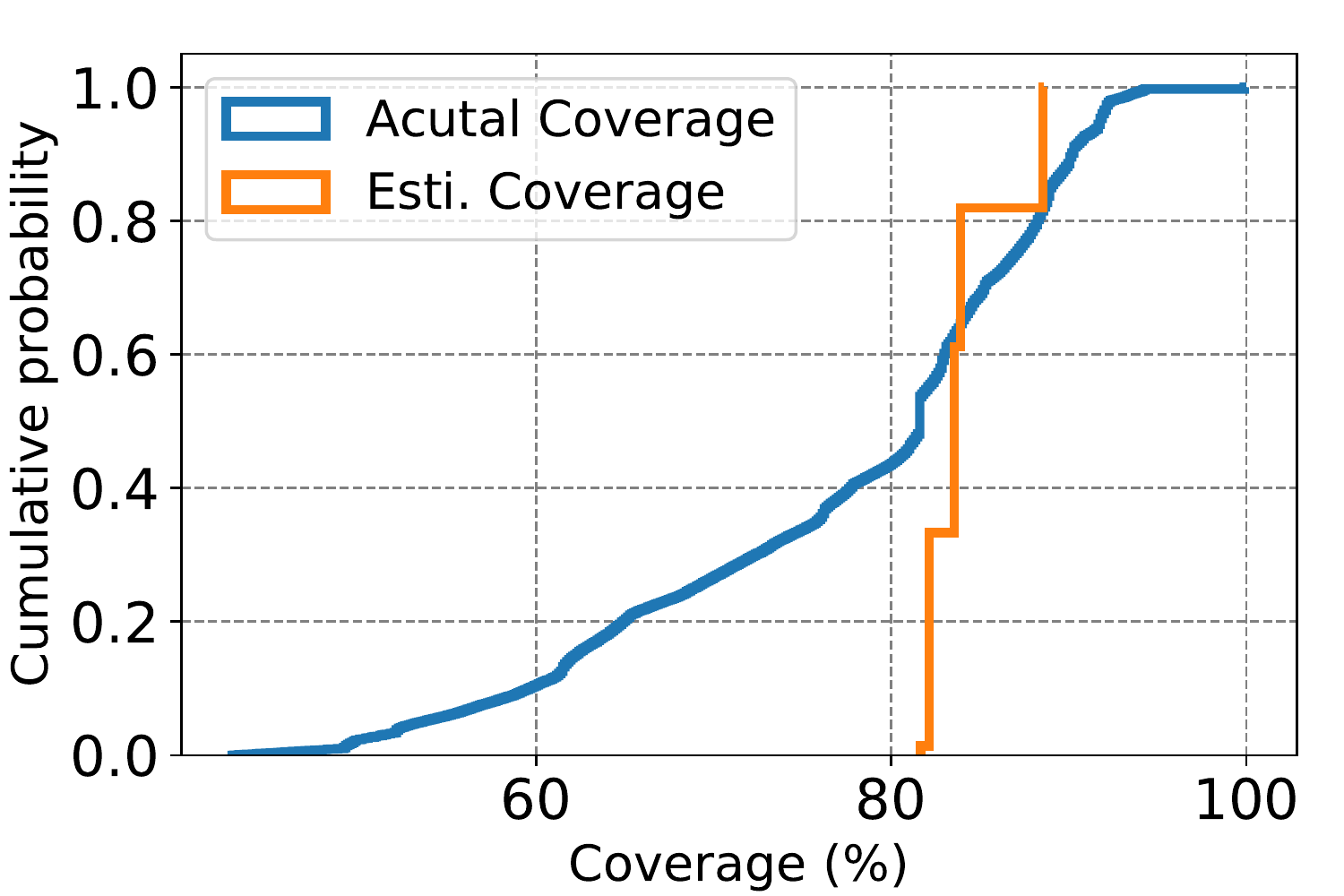}
    	\caption{\small Actual coverage in \emph{LiveMap}.}
    	\label{fig:simulation_coverage_time}
    \end{minipage}
\hfill
    \begin{minipage}[!t]{0.32\linewidth}
        \centering
    	\includegraphics[width=2.3in]{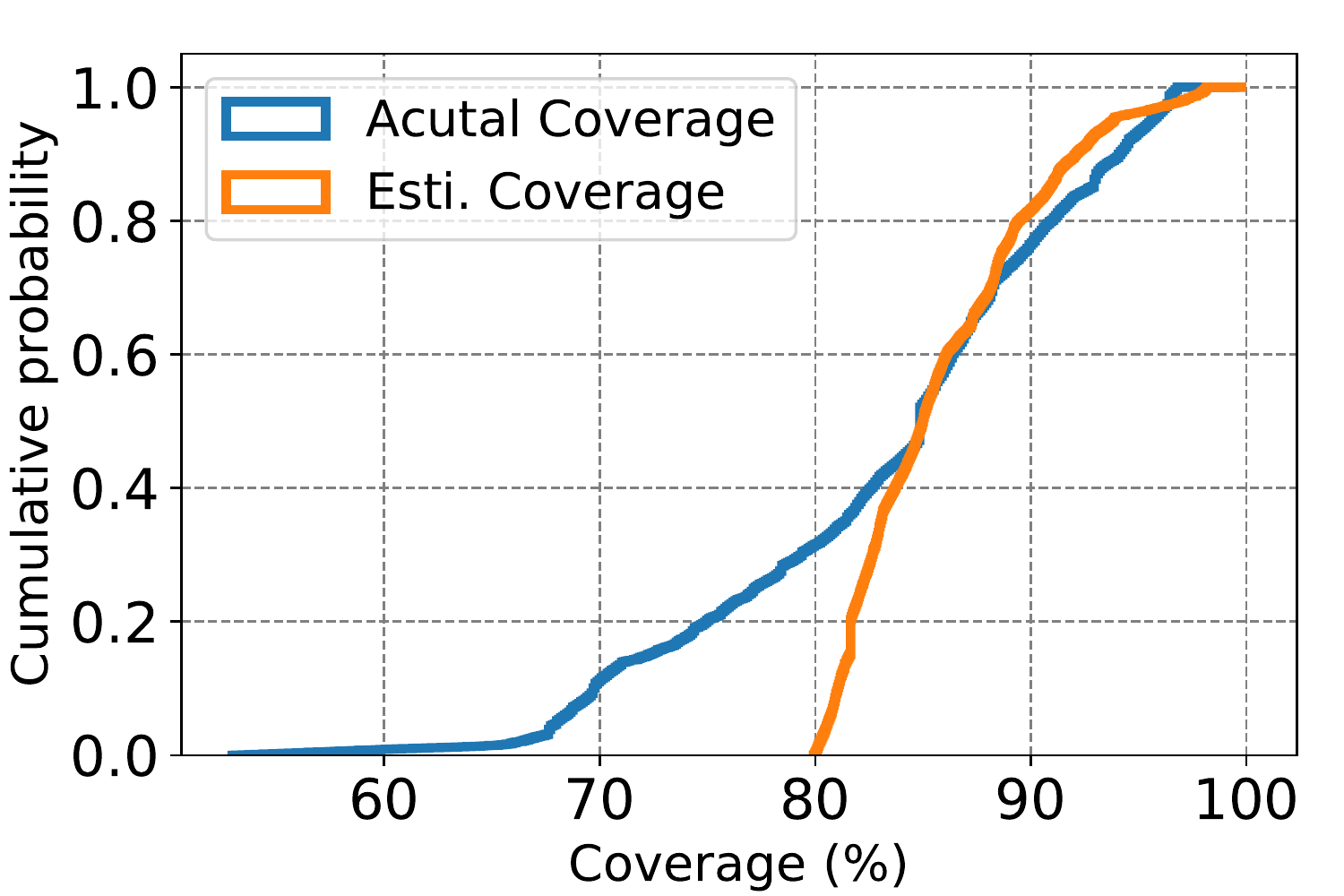}
    	\caption{\small Actual coverage in \emph{LiveMap-Dist}.}
    	\label{fig:simulation_coverage_range}
    \end{minipage}
\hfill
    \begin{minipage}[!t]{0.32\linewidth}
        \centering
    	\includegraphics[width=2.3in]{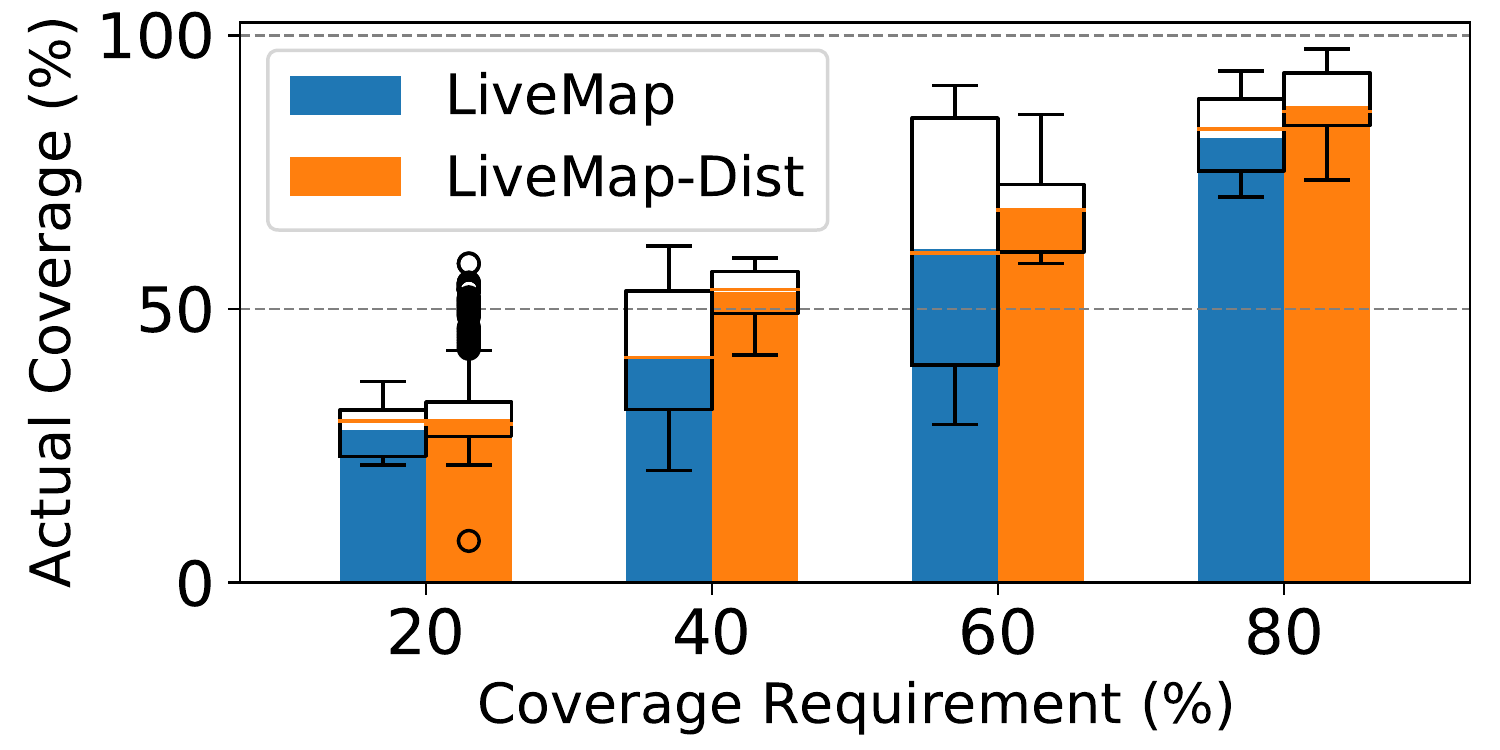}
    	\caption{\small The fulfillment of coverage requirements.}
    	\label{fig:simulation_coverage_rate}
    \end{minipage}
\end{figure*}

% number of server, radio BW, resource utilization (time domain or CDF), better computer capability of vehicle and server

\vspace{-0.05in}
\subsection{Control Plane Performance}
\label{subsec:simulation}
In this part, we evaluate \emph{LiveMap} and \emph{LiveMap-Dist} via large scale network simulation.
The default wireless bandwidth is 0.1 MHz, and the number of CAVs and edge servers are 50 and 5, respectively.
The default coverage range of vehicles are circles with a 50m radius.

\textbf{Overall Performance.} Fig.~\ref{fig:latency_cdf} shows the cumulative probability of latency achieved by different algorithms.
We see that, \emph{LiveMap} and \emph{LiveMap-Dist} are with almost the same performance and achieve the best latency performance among all algorithms.
This result suggests that, although \emph{LiveMap-Dist} has no central information in individual vehicles, its effective design of distributed decision still obtains comparative performance to \emph{LiveMap}.
In addition, RO may occasionally achieve low latency (e.g., less than 150ms), which indicates that the random decision fails to achieve low latency offloadings.
The performance of EO is out-of-the-axis, since offloading the original sensor data overwhelms the wireless transmission and thus substantially deteriorates the latency performance (i.e., all of them are above 500ms).

% With more vehicles in the system, \emph{LiveMap} can obtain higher latency reduction as compared to other algorithms.
% This is because \emph{LiveMap} can maintain the coverage requirement by scheduling fewer ratio of vehicles when there are more vehicles in the system.
% Besides, Fig.~\ref{fig:simulation_carnum_coverage} (b) shows the average latency and scheduled vehicle ratio achieved by \emph{LiveMap} under different coverage requirements, where \emph{LiveMap} schedules fewer vehicles as the loosen of coverage requirement.
% By sacrificing more map coverage, e.g., from 100\% to 60\%, the scheduled vehicle ratio can be dramatically reduced from 100\% to 15.6\%, and the average latency is thus decreased from 231.7ms to 159.1ms in the meantime.
% These results validate that \emph{LiveMap} is scalable under the different number of CAVs. 

% Next, we evaluate the performance of \emph{LiveMap} and \emph{LiveMap-Dist} under different network configurations.
\textbf{Network Dynamics.} 
Fig.~\ref{fig:simulation_carnum_coverage_1} shows the average latency of all algorithms under different number of CAVs.
With 50 vehicles in the system, \emph{LiveMap} and \emph{LiveMap-Dist} can reduce more than 10\% average latency than RM.
When there are more vehicles, \emph{LiveMap} and \emph{LiveMap-Dist} achieve higher reductions of the average latency over the other algorithms.
This is because, when there are more vehicles in the given geographic area, the average AoR of all vehicles are higher in general.
Thus, the efficient design of vehicle scheduling in \emph{LiveMap} and \emph{LiveMap-Dist} can satisfy the coverage requirement by scheduling fewer vehicles.
Fig.~\ref{fig:simulation_bandwidth} and Fig.~\ref{fig:simulation_num_server} show the latency performance under different radio bandwidth and the number of edge servers.
It can be observed that the higher radio bandwidth improves the latency performance, but the improvement gains diminish.
This situation also applies to the number of edge servers. 
In the edge computation model in the network simulator, the more edge servers will mostly reduce the queuing latency of incoming tasks, rather than accelerating their computation.
Thus, as compared to increasing the number of edge servers, improving the capability of individual edge servers may be the better option to reduce the computation latency of tasks.
From these results, we can see \emph{LiveMap} and \emph{LiveMap-Dist} obtain similar latency performance, which validates the effectiveness of the D-HEAD algorithm in \emph{LiveMap-Dist} on optimizing scheduling and offloading for distributed vehicles.

\textbf{Coverage.} Fig.~\ref{fig:simulation_coverage_time} and Fig.~\ref{fig:simulation_coverage_range} show the coverage performance of \emph{LiveMap} and \emph{LiveMap-Dist}.
As \emph{LiveMap} periodically optimizes the vehicle scheduling on a larger time scale, the instant overall coverages may be varying due to the mobility of vehicles during scheduling intervals. 
Fig.~\ref{fig:simulation_coverage_time} shows that the actual instant coverage of \emph{LiveMap} ranges from 55\% to 100\%.
Although the average coverage of 81\% is above the given threshold, instant coverage cannot be guaranteed all the time.
Fig.~\ref{fig:simulation_coverage_range} shows the similar situation in \emph{LiveMap-Dist}, where we simulate the random coverage ranges of vehicles between 25m to 75m radius.
In other words, although \emph{LiveMap-Dist} maintains the coverage requirement asynchronously, its actual instant coverages may be compromised if the coverage ranges of vehicles are very different. 
Fortunately, the coverage ranges of vehicles are relatively static and can be obtained from the global map when they are connected to \emph{LiveMap}. 
Besides, the coverage fulfillment are shown in Fig.~\ref{fig:simulation_coverage_rate}.
We can see that both \emph{LiveMap} and \emph{LiveMap-Dist} can satisfy the coverage requirements on average, and may occasionally violate the coverage requirements in the meantime.
In addition, \emph{LiveMap-Dist} achieves better coverage performance as compared to \emph{LiveMap}, which can be attributed to the asynchronous vehicle scheduling mechanism.

\begin{figure*}[!t]
	\centering
	\includegraphics[width=7.0in]{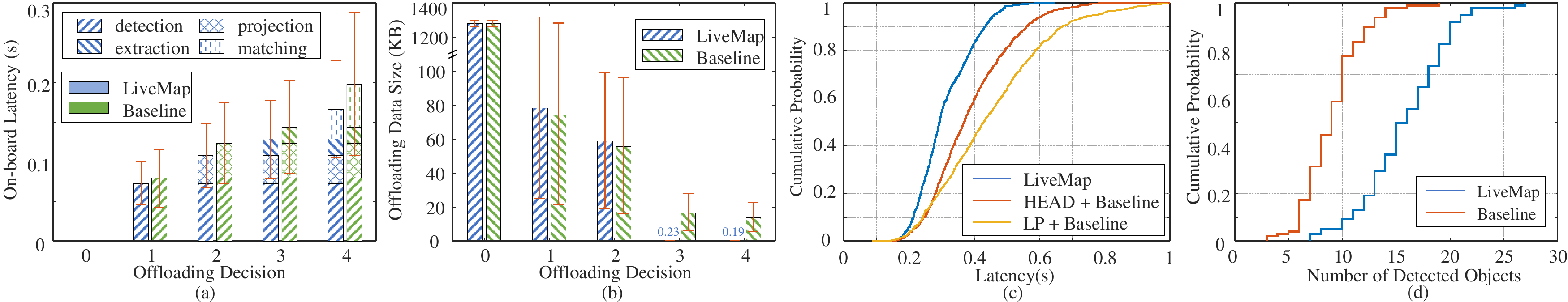}
	\vspace{-0.05in} \caption{\small The system comparison between \emph{LiveMap} and baseline~\cite{liu2021livemap}.}
	\label{fig:system_comparison}
\end{figure*}

\vspace{-0.05in}
\subsection{Data Plane Performance}

We evaluate the data plane of \emph{LiveMap} in terms of processing latency, offloading data size, and overall detection accuracy by comparing it with a baseline.
Note that \emph{LiveMap} and \emph{LiveMap-Dist} use the same data plane.
We build the baseline system by using tiny YOLOv3 model~\cite{redmon2018yolov3} for the object detection component, ORB algorithm~\cite{rublee2011orb} for the feature extraction component, and brutal-force feature matching algorithm.
The other components, e.g., projection and combination, are implemented as same as that of \emph{LiveMap}.

Fig.~\ref{fig:system_comparison} (a) and (b) show the onboard computation latency and intermediate data size under different offloading decisions.
As compared to the baseline system, \emph{LiveMap} obtains lower computation latency in terms of both the mean value and the variance. 
This performance improvement can be attributed to a variety of optimized computation components in \emph{LiveMap}.
For example, we design the object detector to reduce its inference time while achieving a similar accuracy performance in \emph{LiveMap}.
We design the object matcher to use both geo-location and feature distance, which decreases the size of candidate matching objects.
Specifically, the detection in \emph{LiveMap} obtains almost 10\% latency reduction than that of the baseline system.
Besides, we design the feature extractor in \emph{LiveMap} to adopt new autoencoder architecture, which generates condensed but effective features for small cropped images.
Hence, we see a substantially decrease in feature sizes as compared to the baseline system after feature extraction in Fig.~\ref{fig:system_comparison} (b).

% This is because the image preparation (e.g., reading and formatting) and the post-processing (e.g., memory copying from GPU to CPU) account for considerable latency in the embedded GPU platform.
% \emph{LiveMap} can be further integrated with pipelined image operations, which is expected to achieve better latency performance. 
% Besides, because the autoencoder in \emph{LiveMap} generates concise object features, its offloading data size is significantly smaller than that of the baseline system after feature extraction in Fig.~\ref{fig:system_comparison} (b).
% Here, the large variations of offloading data size in the object projection and feature extraction, i.e., offloading decisions 1 and 2, are from the varying number of objects in RGB images. 

Fig.~\ref{fig:system_comparison} (c) shows the CDF of latency achieved by different systems.
Here, we introduce the HEAD algorithm under the baseline system to evaluate the effectiveness of the data plane, as \emph{LiveMap} also uses the HEAD algorithm.
We observe that \emph{LiveMap} decreases 20.1\% and 34.1\% latency on average as compared to the HEAD and LP algorithm under the baseline system, respectively.
In other words, this improvement in latency performance justifies the efficacy of the data plane in \emph{LiveMap}.
In addition, the performance difference between the HEAD algorithm and LP algorithm also validates the efficacy of the control plane of \emph{LiveMap}.
Besides, we evaluate the detection accuracy of \emph{LiveMap} and the baseline system in Fig.~\ref{fig:system_comparison} (d).
Here, the criteria for determining if an object is detected successfully is that: 1) the object \emph{id} should be matched correctly; 2) meanwhile, the deviation of estimated geo-location should be less than 1 meter.
It can be seen that \emph{LiveMap} improves 74.9\% number of detected objects on average as compared to the baseline system, which is mainly attributed to the efficacy of the autoencoder-base feature extractor on the small object images.
Hence, we conclude that the data plane in \emph{LiveMap} outperforms existing solutions in multiple aspects.

% \begin{figure}[!t]
% 	\centering
% 	\includegraphics[width=3.48in]{./fig/traffic_demand_awareness.pdf}
% 	\caption{\small a) system resource usage vs. average traffic within an episode; b) resource usage of network slices.}
% 	\label{fig:traffic_demand_awareness}
% \end{figure}

%%%%%%%%%%%%%%%%% different coverage range, see D-HEAD, and coverage in time domain, insights for periodical scheduling %%%%%%%%%%%%%%%%%%%%
% \hfill
%     \begin{minipage}[!t]{0.32\linewidth}
%         \centering
%     	\includegraphics[width=2.3in]{./fig/simulation_capacity.pdf}
%     	\caption{\small Latency under different server capacities.}
%     	\label{fig:simulation_capacity}
%     \end{minipage}

% \vspace{-0.05in}
\section{Related Work}

% This work is related to computation offloading, resource management, and vehicle perception in mobile networks and edge computing. 

\textbf{Computation offloading} aims to accelerate the computation of mobile devices via offloading the needed data to edge and cloud computing. 
Ran \emph{et. al.}~\cite{ran2018deepdecision} proposed DeepDecision, a new framework to optimize the offloading strategy for augmented reality (AR), while balancing the detection accuracy, video quality, battery, and network data usage.
DARE~\cite{liu2018dare} achieves a dynamic adaptive offloading scheme in mobile augmented reality, which optimizes the offloading image quality and computation models in edge servers according to the availability of network resources.
Liu \emph{et. al.}~\cite{liu2019edge} designed an edge analytics system including parallel offloading and rendering pipeline and object tracking method, which achieves a high-fps and accurate object detection on over-the-shelf AR devices.
These works have shown substantial computation acceleration in mobile edge computing, which inspires the idea of crowdsourcing from CAVs in \emph{LiveMap}.

\textbf{Machine learning} has demonstrated a great potential to handle complex network systems in recent years.
Bao \emph{et. al.}~\cite{bao2019deep} developed Harmony, an ML cluster scheduler (based on DRL techniques), to optimize the placement of ML tasks and accelerate their completion time.
%
% EdgeSlice~\cite{liu2020edgeslice} introduces a decentralized DRL-based algorithm to manage multiple networking and computing resources for satisfying the service level agreement (SLA) of network slices.
%
Wang \emph{et. al.}~\cite{wang2019intelligent} designed DeepCast, which relies on a new DRL algorithm to learn the individualize QoE of online viewers and determine the scheduling and transcoding selection in interactive crowdsourcing livecast.
%
% Mao \emph{et. al.}~\cite{mao2016resource} designed DeepRM tothat optimizes the admission control and resource allocation of users, which obtains a promising reduction on the average slowdown of user tasks as compared to heuristic solutions.
%
% Xu \emph{et. al.}~\cite{xu2018experience} uses a DDPG-based algorithm to solve the traffic engineering problem, i.e., allocating the bandwidth of network links, and achieves significant latency reduction without prior performance functions.
%
% Bega \emph{et. al.}~\cite{bega2019machine} proposed an N3AC algorithm that optimizes the admission control of slices with the DRL technique to maximize the revenue of the orchestrator while meeting the SLA of admitted slices.
%
Existing works, however, tackle unconstrained DRL problems, whose action and state space are fixed. 
In contrast, \emph{LiveMap} deals with the problem under varying number of CAVs in both centralized and distributed manner.

\textbf{Vehicle co-perception} has shown promising accuracy performance in autonomous driving, which aggregates multi-viewed sensor data from different vehicles.
Qiu \emph{et. al.} developed augmented vehicular reality (AVR)~\cite{qiu2018avr} and AutoCast~\cite{qiu2021autocast} to achieve infrastructure-less cooperative perception using direct vehicle-to-vehicle communication.
Different mechanisms are applied to reduce the size of transmission data, e.g., distinguishing dynamic objects from static objects and prioritizing safety-critical transmissions.
Ahmad \emph{et. al.}~\cite{ahmad2020carmap} designed CarMap, a feature-represented lean 3D map via crowdsourcing sensor data from CAVs.
This map achieves near real-time update, which is accomplished by excluding transient information, e.g., pedestrians, from map generation and processing.
These systems mainly share static information (e.g., point clouds among vehicles), while \emph{LiveMap} enables dynamic information sharing (e.g., features or images) according to contextual offloading decisions of vehicles.

\vspace{-0.05in}
\section{Conclusion}
In this work, we presented a new real-time dynamic map that achieves efficient perception sharing among crowdsourcing vehicles.
We designed an efficient data plane to detect, match, and track objects on the road in the time scale of subseconds.
We designed an intelligent control plane with two new algorithms to schedule vehicles and optimize offloading decisions under network dynamics. 
We have shown our solution achieves better latency, coverage, and accuracy performance than existing solutions through both experiments and simulations.

\bibliographystyle{IEEEtran}
\bibliography{ref/reference, ref/qiang}

\ifCLASSOPTIONcaptionsoff
  \newpage
\fi



%\bibliographystyle{IEEEtran}


\vspace{-0.3in}
\begin{IEEEbiography}[{\includegraphics[width=1in,height=1.25in,clip,keepaspectratio]{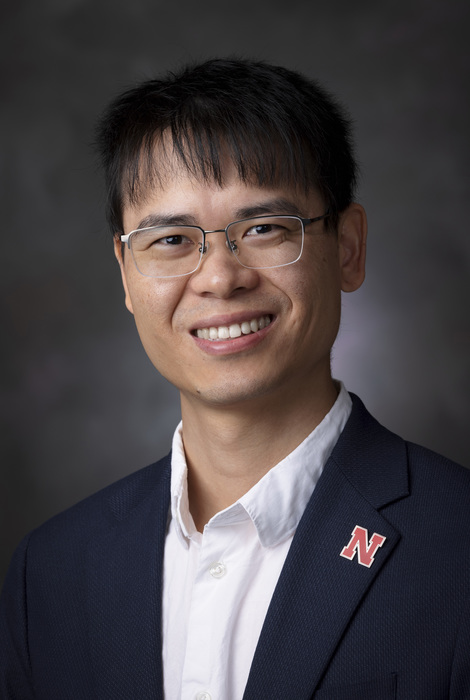}}]%
{Qiang Liu} is currently an Assistant Professor at the School of Computing, University of Nebraska-Lincoln. He earned his Ph.D. degree in Electrical Engineering from the University of North Carolina at Charlotte (UNCC) in 2020. His paper won IEEE International Conference on Communications (ICC) Best Paper Award 2019, 2022, and IEEE Communications Society's Transmission, Access, and Optical Systems (TAOS) Best Paper Award 2019. His research interests lie in the broad field of wireless communication, computer networking, edge computing, and machine learning.
\end{IEEEbiography}
\vspace{-0.3in}
\begin{IEEEbiography}[{\includegraphics[width=1in,height=1.25in,clip,keepaspectratio]{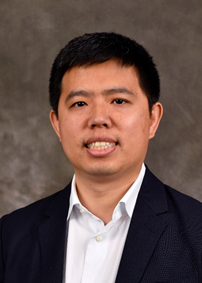}}]%
{Tao Han} (M’15-SM’20) is an Associate Professor in the Department of Electrical and Computer Engineering at New Jersey Institute of Technology (NJIT) and an IEEE Senior Member. Before joining NJIT, Dr. Han was an Assistant Professor in the Department of Electrical and Computer Engineering at the University of North Carolina at Charlotte. Dr. Han received his Ph.D. in Electrical Engineering from NJIT in 2015 and is the recipient of NSF CAREER Award 2021, Newark College of Engineering Outstanding Dissertation Award 2016, NJIT Hashimoto Prize 2015, and New Jersey Inventors Hall of Fame Graduate Student Award 2014. His papers win IEEE International Conference on Communications (ICC) Best Paper Award 2019 and IEEE Communications Society’s Transmission, Access, and Optical Systems (TAOS) Best Paper Award 2019. His research interest includes mobile edge computing, machine learning, mobile X reality, 5G system, Internet of Things, and smart grid.
\end{IEEEbiography}

%\newpage
\vspace{-0.3in}
\begin{IEEEbiography}[{\includegraphics[width=1in,height=1.25in,clip,keepaspectratio]{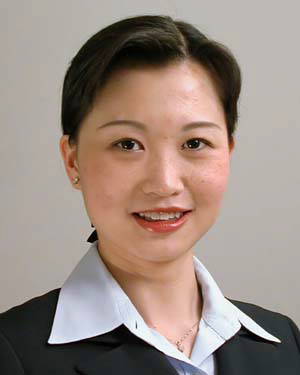}}]%
{Jiang Xie} (Fellow, IEEE) received the B.E. degree from Tsinghua University, Beijing, China, the M.Phil. degree from the Hong Kong University of Science and Technology, and the M.S. and Ph.D. degrees from Georgia Institute of Technology, all in electrical and computer engineering. She joined the Department of Electrical and Computer Engineering, the University of North Carolina at Charlotte (UNC Charlotte) as an Assistant Professor in August 2004, where she is currently a Full Professor. Her current research interests include resource and mobility management in wireless networks, mobile computing, Internet of Things, cloud/edge computing, and virtual/augmented reality. She received the U.S. National Science Foundation NSF Faculty Early Career Development (CAREER) Award in 2010, the Best Paper Award from IEEE Global Communications Conference in 2017, the Best Paper Award from IEEE/WIC/ACM International Conference on Intelligent Agent Technology in 2010, and the Graduate Teaching Excellence Award from the College of Engineering at UNC-Charlotte in 2007. She is on the editorial boards of the IEEE Transactions on Wireless Communications, IEEE Transactions on Sustainable Computing, and Journal of Network and Computer Applications (Elsevier). She is a Senior Member of ACM. 
\end{IEEEbiography}

\vspace{-0.3in}
\begin{IEEEbiography}[{\includegraphics[width=1in,height=1.25in,clip,keepaspectratio]{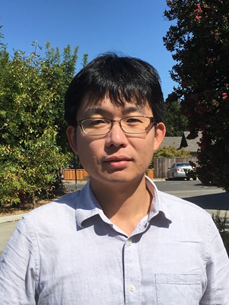}}]%
{BaekGyu Kim}
earned B.S. and M.S. in electrical engineering and computer science from Kyungpook National University in South Korea, and earned Ph.D in computer science from University of Pennsylvania. He is currently an assistant professor at Department of Information and Communication Engineering in DGIST (Daegu Gyeongbuk Institute of Science and Technology). Before joining DGIST, he was a principal researcher at Toyota Motor North America, InfoTech Labs where he conducted industrial research on connected car software platforms for six years. His primary research area includes verification, validation and optimization techniques to guarantee assurances for Internet of Things and Cyber Physical Systems.
\end{IEEEbiography}

%\begin{IEEEbiographynophoto}{Ignacio Ramos}
%(S'12) received the B.S. degree in electrical engineering from the University of Illinois at Chicago in 2009, and is currently working toward the Ph.D. degree at the University of Colorado at Boulder. From 2009 to 2011, he was with the Power and Electronic Systems Department at Raytheon IDS, Sudbury, MA. His research interests include high-efficiency microwave power amplifiers, microwave DC/DC converters, radar systems, and wireless power transmission.

%% insert where needed to balance the two columns on the last page with
%%\newpage

% ==== SWITCH OFF the BIO for submission
% ==== SWITCH OFF the BIO for submission


\vfill


\end{document}